# p-(001)NiO/n-(0001)ZnO Heterostructures based Ultraviolet Photodetectors

Amandeep Kaur, Bhabani Prasad Sahu, Ajoy Biswas, Subhabrata Dhar*

Department of Physics, Indian Institute of Technology Bombay, Powai, Mumbai 400076, India

*Email: dhar@phy.iitb.ac.in

**ABSTRACT:**

We investigate the potential of epitaxial (001)p-NiO/(0001)n-ZnO heterostructures grown on (0001)sapphire substrates by pulsed laser deposition technique for ultraviolet photodetector application. Our study reveals that in the self-powered mode, these devices can serve as effective photodetectors for the UV-A band (320–400 nm) with response time as short as ~400μs. Peak responsivity as high as 5mA/W at zero bias condition have been achieved. These devices also show a very high level of stability under repeated on/off illumination cycles over a long period of time. Furthermore, we find that the response time of these detectors can be controlled from several microseconds to thousands of seconds by applying bias both in the forward and the reverse directions. This persistent photoconductivity effect has been explained in terms of the field induced change in the capture barrier height associated with certain traps located at the junction.

## INTRODUCTION:

Ultraviolet (UV) component in the sun light can be quite harmful as it can cause skin cancer, damage our immune system and accelerate the aging process [1–3]. The current scenario of gradual depletion of the UV protective Ozone layer in our atmosphere has enhanced the need for developing efficient and cost-effective UV detectors, which can also be used for various other applications such as biological analysis, flame detection, UV imaging, pollution monitoring, air/water purification, submarine oil leakage monitoring and space communications [4–8]. Photodetectors are mostly supported by the battery power. This is not an environment friendly solution because the toxic chemicals used in the batteries may pose health hazards, if these are not properly recycled [9]. Self-powered photodiodes have thus garnered significant attention in the recent years [10,11]. These photodiodes harness the photovoltaic (PV) effect to generate power [12]. Out of the two variants of this class of detectors, p-n junctions are preferred over the Schottky junctions as the former type of devices show comparatively lower dark current and response time [13,14]

Commercially available photodetectors are silicon-based, which has an optical bandgap of only 1.1 eV. This means that the devices require expensive optical filters for blocking the visible lights to function effectively in the UV regime [7,15]. Moreover, these devices show significant leakage current, poor efficiency, and reduced sensitivity for the detection of the UV signals. Wide bandgap materials are the natural choices for developing the UV photodetectors with higher efficiency [16–19]. Among them, nickel oxide (NiO) stands out as a compelling candidate for UV detection due to its wide indirect bandgap (3.2–3.6 eV) and superior environmental stability. Moreover, abundance of NiO in the earth crust makes it a cost-effective and environmentally friendly material. NiO with NaCl lattice structure is one of the few oxide semiconductors, where stable p-type doping is possible. As grown material is often found to be unintentionally p-type doped. This is attributed to nickel vacancies, which are believed to act as shallow acceptors in the material [20]. On the other hand, Zinc oxide (ZnO) is a direct wide bandgap (~3.40 eV) semiconductor with a very high exciton binding energy of 60 meV [21]. ZnO can easily be doped n-type, while p-type doping has been found to be challenging in this semiconductor. The material is also environmentally friendly and chemically stable. ZnO has a stable hexagonal wurtzite phase [22,23], (0001) plane of which, offers similar lattice symmetry as that of (111) plane of NiO. All these brighten the prospect of growing p-NiO/n-ZnO epitaxial heterojunction for UV photodetectors.

There are indeed several reports of p-NiO/n-ZnO heterojunctions-based photodetectors, where the junctions are made of polycrystalline films deposited on polymer or glass substrates [24–33]. Since polycrystalline films are accompanied by numerous structural defects and imperfections, the efficiency and reproducibility of the devices are unlikely to be up to the mark. p-NiO/n-ZnO epitaxial heterojunctions are expected to perform as more reliable and efficient UV photodetectors. However, there are hardly any report of such devices in the literature.

In this study, epitaxial heterostructures of p-(001)NiO/n-(0001)ZnO on (0001) sapphire substrates are grown using pulsed laser deposition (PLD) technique. We have investigated the responsivity and time response characteristics of these heterostructures by subjecting them to photon pulses of varying energies. Our findings indicate that even in self-powered mode, these devices can function as visible-blind ultraviolet photodetectors with impressive responsivity and speed. Notably, peak responsivity values as high as 5mA/W at zero bias condition were achieved in these devices. Moreover, the rise and decay times of such devices were measured to be 444/628μs, respectively. The devices are found to show persistent photoconductivity effect both under forward and reverse bias conditions. Interestingly, the photoconductivity response time can be varied from several microseconds to thousands of seconds by applying bias. This has been explained in terms of the field induced change in the capture barrier height associated with certain traps located at the junction. Note that such electrically controllable PPC behaviour can be utilized for optoelectronic artificial synaptic devices [34].

**EXPERIMENTAL DETAILS:**

Pulsed Laser Deposition (PLD) technique was used to grow the NiO/ZnO heterostructure on *c*-sapphire substrates. Base pressure of the growth chamber was maintained at less than $1 \times 10^{-5}$ mbar. NiO and ZnO pellets were ablated by an excimer KrF laser with a wavelength of 248 nm and a pulse width of 25 ns. The energy density and frequency of the laser pulse was 2.0 J/cm$^2$ and 5 Hz, respectively. The ZnO layer was first grown at a substrate temperature $T_G = 500°C$ and oxygen pressures $\emptyset_{O_2}$ of $2 \times 10^{-2}$ mbar for 10,000 pulse counts. NiO layer was subsequently grown on top of the ZnO film at $T_G = 400°C$ and $\emptyset_{O_2} = 2 \times 10^{-2}$ mbar for 20,000 counts. During the growth, a portion of the ZnO/sapphire template was covered with a piece of sapphire to keep that part free from NiO deposition. More details of the growth process can be found elsewhere [35]. X-ray diffraction (XRD) studies were conducted on the sample using a Rigaku Smart Lab high-resolution XRD system. Surface morphology and roughness were studied using atomic force microscopy (AFM) and scanning electron microscopy (SEM). To investigate the interface quality of the grown heterostructure, a lamella was prepared using focused ion beam (FIB) lift-out techniques in a Helios 5 UC Thermo-scientific dual-beam instrument with a resolution ~0.6 nm. Cross-sectional High-Resolution TEM (HRTEM) investigations were carried out on this lamella using Thermo Scientific, Themis 300 G3, 300kV high-resolution transmission microscope with point resolution of 0.2 nm. Circular metal contacts of Ti/Au (20nm/80nm) and Ni/Au (20nm/80nm) each with diameter 1mm were deposited using thermal evaporation technique on the ZnO and NiO sides, respectively. The distance between these contacts was kept to be ~4mm. The current-voltage characteristics between the metal contacts on each side show more or less linear profiles suggesting Ohmic behaviour (see supplementary information S1). I-V measurements were carried out using Keithley 6487 pico-ammeter voltage source. For photoconductivity measurements, a 150-W Xenon white-light source was used. After passing through a monochromator to select the photon energy the incident beam was focused onto the device. The intensity of the light falling on the active area of the device was measured using a power meter. To record the temporal response, the device was illuminated with monochromatic lights, whose intensity was modulated with a chopper. Resulting photocurrent signal was then analysed using an oscilloscope with a bandwidth of 2.5 GHz.

**RESULTS AND DISCUSSIONS:**

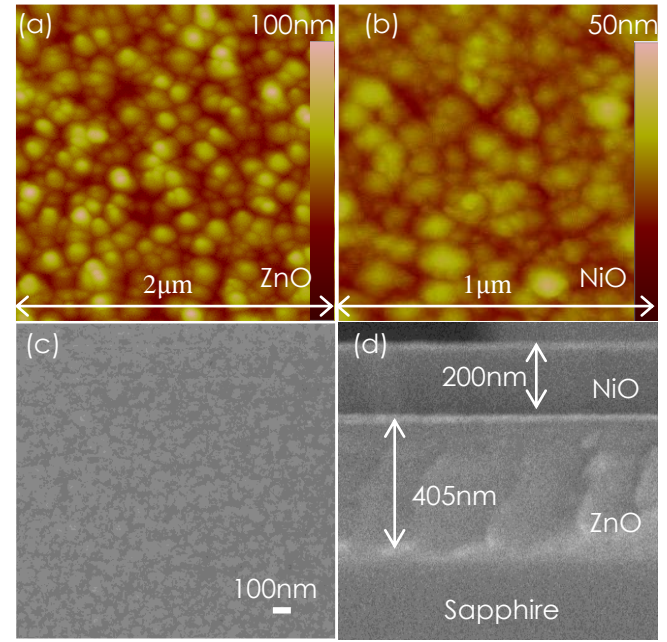

Fig. 1 AFM images recorded on (a) the bare surface of ZnO and (b) the surface of the NiO film grown on top of the ZnO layer. (c) SEM micrograph of the NiO surface. (d) cross-sectional SEM image of the heterostructure.

Figures 1(a) and (b) present the AFM surface images of the grown ZnO and NiO layers, respectively, which reveal smooth and continuous morphology in both the cases. The rms roughness of the ZnO and NiO surfaces are estimated to be 9.82 nm and 3.74 nm, respectively. Continuous and smooth deposition of NiO is further confirmed by the SEM surface micrographs of the film as shown in Fig. 1(c). The cross-sectional SEM image of the heterojunction is presented in Fig. 1(d). The cross-sectional image shows the deposition of ~ 200 and 400 nm thick NiO and ZnO films, respectively.

Figure 2(a) displays the $\omega - 2\theta$ X-ray diffraction (XRD) scan recorded for the sample. Apart from sapphire (0006), NiO (002), and ZnO (0002) peaks and their higher order reflections no other features could be found in the scan. This suggests growth of (001)NiO layer on top of (0001) ZnO film. Fig. 2(b) shows the wide angle $\phi$- scan recorded for $(10\bar{1}0)$ ZnO [blue] and $(200)$NiO [pink] reflections of the sample. In case of ZnO, six equidistant peaks display the hexagonal symmetry of the lattice. While the observation of twelve equidistant peaks, in case of NiO, demonstrates the co-existence of three 30°- rotated cubic domains of (001)NiO on (0001)ZnO surface [35]. All these studies confirm the epitaxial nature of the grown structure.

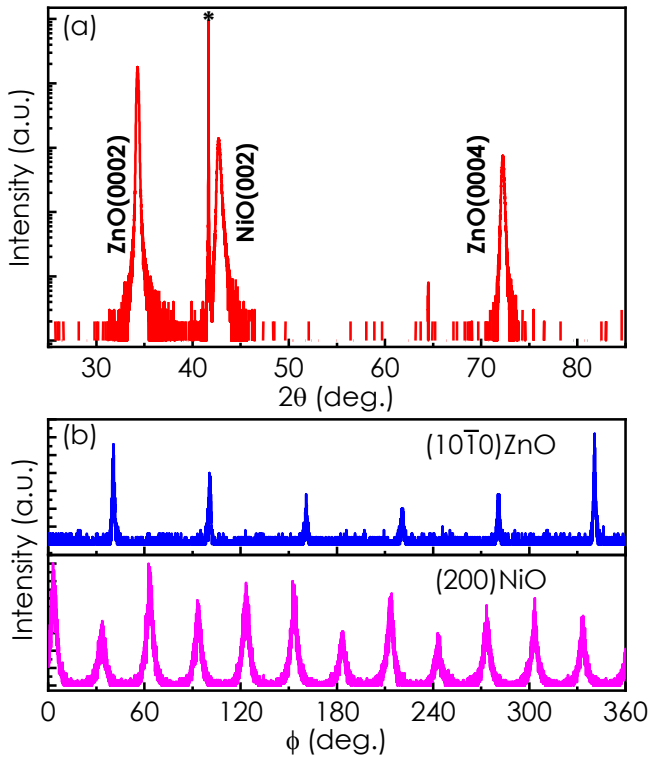

Fig.2 (a) ω − 2θ X-ray diffraction (XRD) scan recorded for the p-NiO/n-ZnO heterojunction. The asterisk marks the reflection from the underlying sapphire substrate. (b) In-plane $\phi$- scan recorded for (200)NiO reflection.

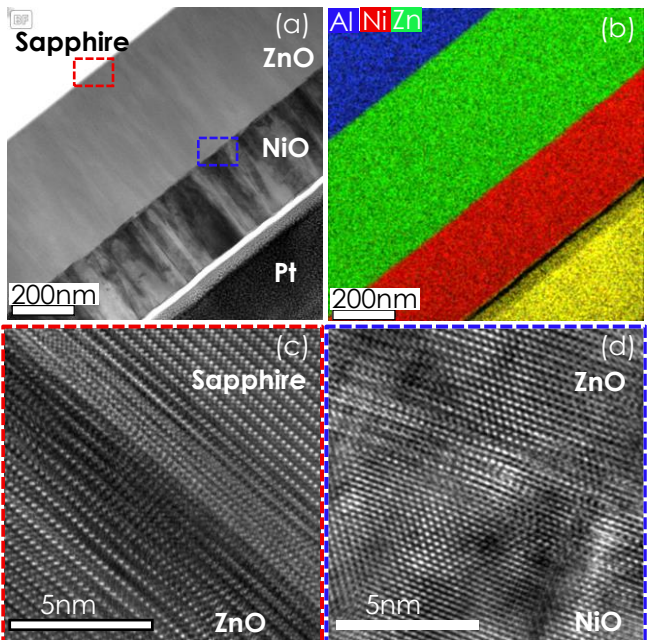

Fig. 3 (a) Bright field image of the lamella prepared from the sample. (b) EDS map of the lamella showing individual layers. High-resolution image of the (c) ZnO and sapphire interface (d) ZnO and NiO interface.

Figure 3(a) shows the cross-sectional transmission electron microscopy (TEM) image of the lamella made from the sample. A sharp material contrast is evident in Fig. 3(b) from EDS mapping at the junction, suggesting an abrupt interface. Figures 3(c) and 3(d) present the high-resolution transmission electron microscopy (HRTEM) images recorded at the ZnO-sapphire and NiO-ZnO interfaces, respectively. The abruptness of both the interfaces is quite evident from these figures. It is noticeable that the interface related disorder diminishes only within a few nanometres at the ZnO/NiO interface suggesting a very sharp boundary.

Current-voltage (I-V) profiles are measured between the NiO and the ZnO contact pads under dark and illumination (with light of wavelength 366 nm) conditions at room temperature. The data are plotted in Fig. 4(a). Evidently, both in the dark and the illumination conditions the I-V profiles show rectifying characteristics implying the presence of a p-n junction. This can be attributed to the unintentional p(n)-type doping of NiO (ZnO) film during growth. It is noteworthy that as grown NiO (ZnO) films grown by this technique are often found to exhibit p(n)-type conductivity. Under illumination, the current increases to 40nA at zero bias condition, as shown in the inset of the figure. This clearly demonstrates the photovoltaic effect highlighting the potential of the system to function effectively as a self-powered photodetector.

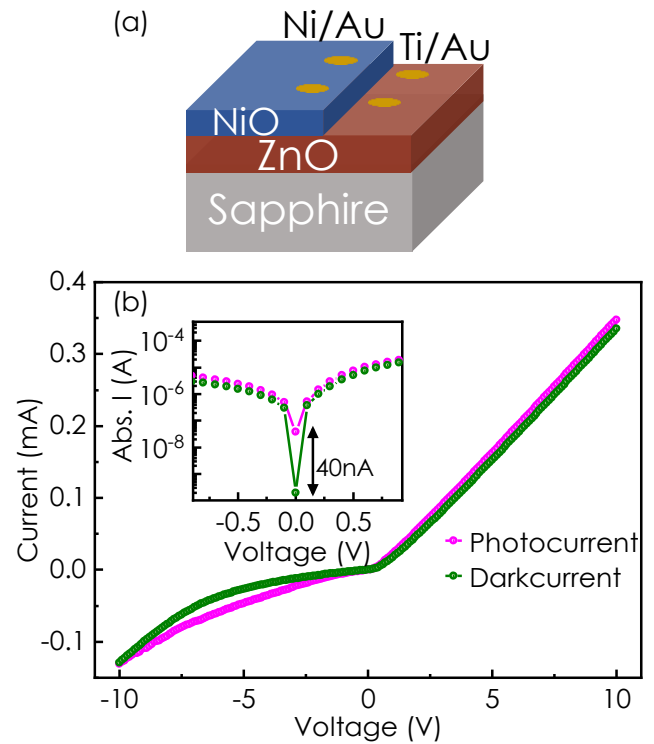

Fig.4(a) Schematic view of the device. (b) I-V characteristics of the p-NiO/n-ZnO heterojunction under dark and illuminated with 366 nm light (with power density of 77μW/cm$^{-2}$) conditions. The inset shows an expanded view of the profiles around zero bias.

Responsivity (R) of a photodiode is defined as $R = (I_\lambda − I_D)/P_i$, where, $I_\lambda$ the photocurrent at a specific wavelength (λ), $I_D$ denotes the dark current, and $P_i$ indicates the optical power delivered to the effective active area (A) of the photodiode [36]. Figure 5 depicts the responsivity as a function of wavelength for the heterojunction operating in the self-powered mode. The spectrum displays a peak at ~366 nm. At zero bias, the heterojunction device exhibits a peak responsivity of 5 mA/W. Detectivity, which is given by $D = R\sqrt{A}/\sqrt{2qI_D}$, can be estimated to be $3.6 \times 10^{11}$ Jones at zero bias at the peak.

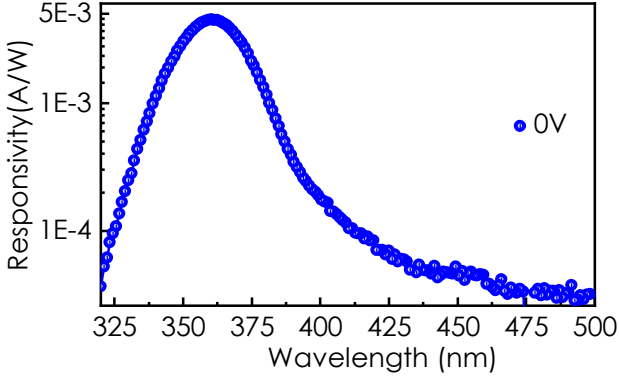

Fig. 5 Spectral responsivity profile of the device at zero bias.

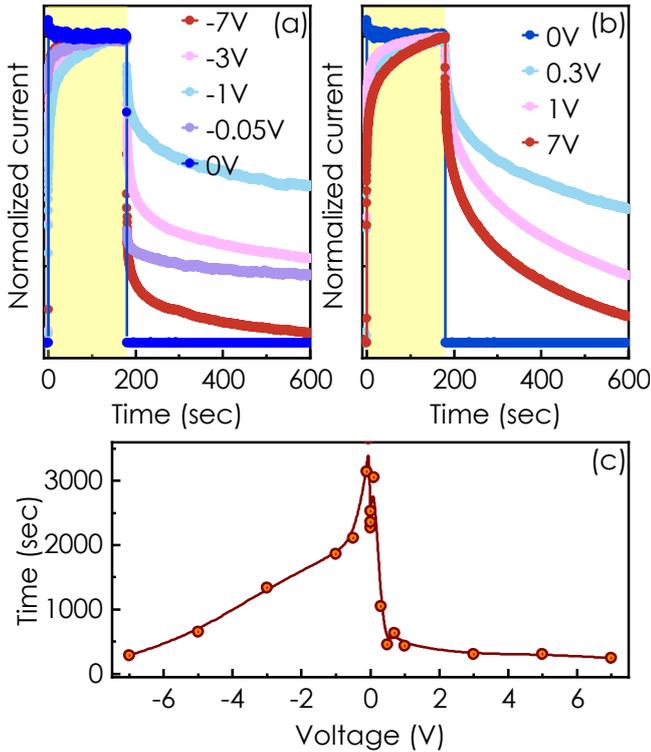

Fig.6 Time-response profiles of the device under different (a) reverse and (b) forward bias conditions after switching on and off the illumination with 366 nm light. Each profile is normalized with respect to the photocurrent maximum after subtracting the dark current. (c) Decay time-constant $\tau_d$ as a function of the applied bias.

Figure 6(a) and (b) plot the normalized time response profiles of the photodiode under the reverse and forward bias conditions as the 366nm illumination with intensity 0.6mW/cm² is switched on and then turned off after 180-seconds of exposure. It is interesting to note that at the zero-bias condition, the photocurrent rapidly goes to the saturation when the light is on and promptly returns to the baseline when it is turned off. However, a small positive or negative bias dramatically enhances both the growth and the decay times. Moreover, the rise and the fall times become bias dependent as can be seen in Fig. 6(c), where the time constant $\tau_d$ for the decay is plotted as a function of the applied voltage. Note that $\tau_d$ is estimated from the slope of the log-of-current versus time plots as shown in the supplementary information S2. Evidently, the time constant decreases from its maximum near $V = 0$ as the bias is increased both in the forward and the reverse directions.

Note that at the zero-bias condition, the current developed in the circuit upon illumination must be arising from the photovoltaic effect at the hetero p-n junction. This current is generated due to the diffusion of minority carriers at the boundaries between the neutral and the depletion regions in p- and/or n-sides [37–39]. When the device is illuminated with 366 nm photons, the light is absorbed only in the ZnO side. The photovoltaic effect is generated only in the ZnO side and it should run a hole-current through the device. When the device is illuminated under the forward(reverse) bias condition, apart from the photovoltaic current, the drift current due to the majority (minority) carriers is set in [40–42]. Flow of different types of currents in various biasing conditions are schematically shown in Fig. 7. The slow photo-response of the device under the forward/reverse biases (as shown in Fig. 6) can be attributed to the persistent photoconductivity (PPC) effect resulting either(both) from the bulk of the two layers or(and) the interface. Since the photo-voltaic current is expected to vanish as soon as the light is switched off, the device shows the fastest photo response at zero-bias condition, when the photocurrent is determined solely by the photovoltaic effect. Under the forward or reverse bias conditions, whenever the drift current starts to dominate over the photovoltaic current, PPC effect governs the time response of the device. The reduction of the decay time with the increase of bias both in the forward and reverse directions [as shown in Fig. 5(c)] may suggest an electric field induced suppression of the capture barrier height that is responsible for the PPC effect. It is plausible that certain traps at the interface offer capture barrier for the photoexcited carriers (either electrons or holes), which is originally asymmetric in nature, as shown in the inset of Fig.7(a) (green colour). When the p-n junction is developed at the interface, the depletion field modifies the shape of the potential. At the zero-bias condition ($V_b = 0$), the barrier height in the NiO side increases, while that is decreased in the ZnO side as shown in the inset of Fig.7(a). As the reverse bias increases, the barrier in the ZnO side gradually suppresses as depicted in the inset of Fig. 7(b). On the other hand, with the increase of the forward bias, the height of the barrier in the NiO side reduces [see the inset of Fig.7(c)]. Therefore, bias applied either in the forward or the reverse directions facilitates the escape of the trapped carriers, which leads to the decrease of the time constant.

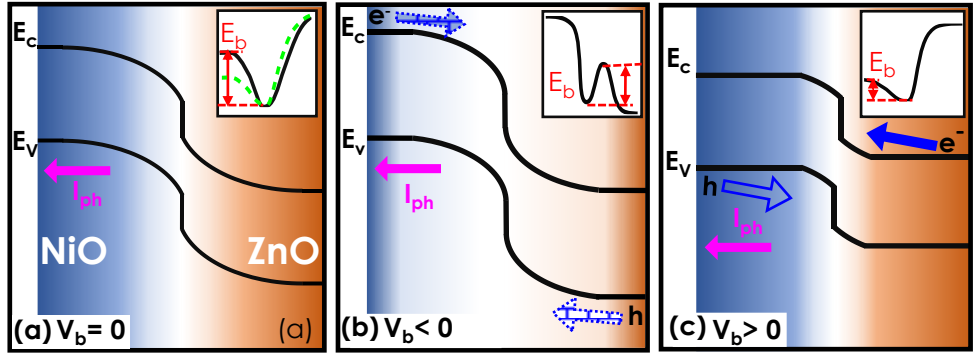

Fig.7 Schematic depiction of the formation of the depletion region as well as the band diagrams under (a) zero, (b) reverse and (c) forward bias conditions at the p-NiO/n-ZnO junction. Drift currents due to majority and minority carriers and the current due to photo-voltaic effect are marked in the respective panels. Insets of panel (a) shows the cartoons of the trap potential in its original form (green colour) and after the modification due to the formation of the pn junction (black colour). Inset of panel (b) and (c) depict schematically the potential under the reverse and forward bias conditions, respectively.

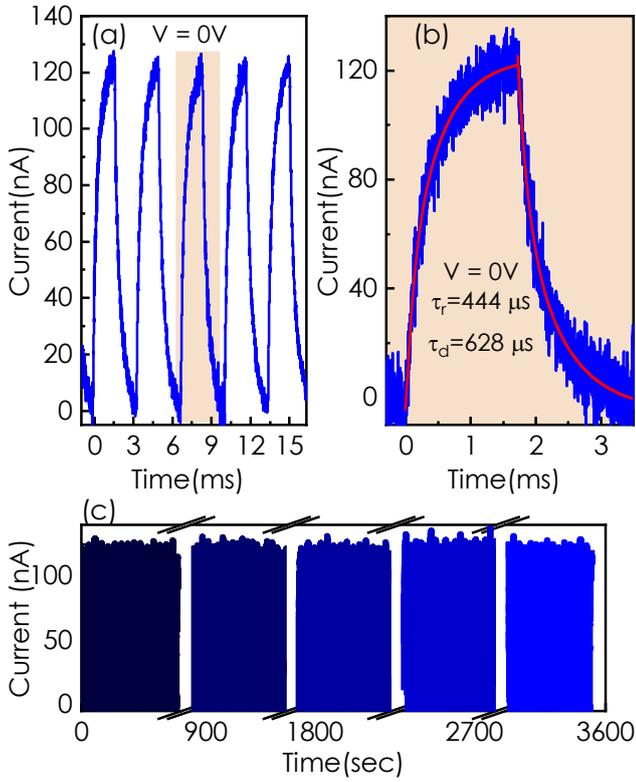

Fig.8: (a) Photocurrent as a function of time recorded under zero-bias as the device is periodically illuminated with a 366 nm monochromatic light (power density of ~0.3mW/cm$^{-2}$) using a chopper. (b) The fitting of the growth and the decay parts of the response. (c) Stability of the photodiode under repeated on/off illumination cycles.

Response of the photodiode under the zero-bias condition as the illumination with 366nm light is modulated using a chopper, is presented in Fig.8(a). The fitting of the rise and decay parts of the response profile with $I(t) = I_{\text{off}} + A\left[1 - \exp\left(-\frac{t-t_0}{\tau_r}\right)\right]$ and $I(t) = I_{\text{off}} + A\left[\exp\left(-\frac{t-t_0}{\tau_d}\right)\right]$, respectively, is shown in Fig.8(b). The rise $(\tau_r)$ and the decay times $(\tau_d)$ are estimated to be 444μs and 628μs, respectively, in this device. Furthermore, the stability of the photodiode under multiple exposures is a crucial parameter for assessing its practical applicability. Fig.8(c) demonstrates the stability of our photodiode under repeated on/off illumination cycles over a period of one hour. Notably, the performance of the photodiode remains consistent and does not degrade over a very long time, indicating a high degree of stability.

Table 1 provides a comparison of the responsivity, detectivity, and response time of our photodetectors with other p-NiO/n-ZnO based PDs reported in the literature. One can see that there are only two reports, which claim faster response time than our device. M. Patel et al. reported a rise time of 41/71μs, but the responsivity was found to be only 20μA/W. On the other hand, A.K. Rana et al. reported a responsivity of 0.29 A/W and response times of 3.92/8.9μs. However, as the authors have pointed out, the origin of the fast photo-response of their devices is the interplay of pyro-phototronic and photovoltaic effects. The photoinduced pyroelectric effect emerges from the generation of heat in the ZnO layer upon UV illumination [43]. This rapid temperature increase within the ZnO layer prompts the distribution of polarized pyroelectric charges to align with the direction of the photovoltaic current, leading to a sharp initial rise in current. However, following this initial spike, the current gradually decreases and stabilize as the illumination persists and the temperature attains a steady state value. Therefore, the steady-state responsivity of these devices is much less than the peak responsivity.

Table 1: Comparison of p-NiO/n-ZnO heterojunctions reported in the literature in terms of responsivity, detectivity, and response times.

| System | Growth technique | Growth Quality | Bias | Responsivity (A/W) | Detectivity (Jones) | Rise /Decay time | Ref. |
|---|---|---|---|---|---|---|---|
| p-NiO/n-ZnO | Rf-sputtered | *ZnO*- polycrystalline *NiO*- polycrystalline | -1.2 | 0.19 | $3.8\times10^{12}$ | 323 /12ms | [24] |
| p-NiO/ZnO | Solution-processed | *ZnO*- Crystalline *NiO*- polycrystalline | 0 | 0.44m | | 0.23 /0.21 s | [25] |
| p-NiO/n-ZnO | Solution method | *ZnO*- polycrystalline *NiO*- polycrystalline | -1 | 0.28 | $6.3\times10^{11}$ | 0.28 /5.4 s | [44] |
| ZnO–NiO | MOCVD | *ZnO*- Crystalline *NiO*- polycrystalline | 0 | 0.493m | | 10/30.3 μs | [26] |
| NiO/ZnO | Hydrothermal method | *ZnO*- polycrystalline *NiO*- polycrystalline | -3 | 2.27 | | -/30sec | [27] |
| NiO/ZnO | Sol–gel processing | *ZnO*- polycrystalline *NiO*- polycrystalline | -5 | 21.8 | $1.6\times10^{12}$ | - | [28] |
| ZnO-NiO | Solution process | *ZnO*- polycrystalline *NiO*- polycrystalline | −1 | 10.2 | $1\times10^{12}$ | 0.2/0.18 s | [29] |
| p-NiO/n-ZnO | Sputtered | *ZnO*- polycrystalline *NiO*- polycrystalline | 0 | 0.29 | $2.75\times10^{11}$ | 3.92/8.9μs | [30] |
| NiO/ZnO | Sputtered | *ZnO*- polycrystalline *NiO*- nanocrystalline | 0 | 20μ | $7.2 \times 10^{11}$ | 41/71μs | [31] |
| n-ZnO/p-NiO | Dc magnetron sputtering | *ZnO*- polycrystalline *NiO*-polycrystalline | 0 | 13.01m | $5.66\times10^{11}$ | 206/477ms | [32] |
| p-NiO/n-ZnO | Dc magnetron sputtering | *ZnO*- Crystalline *NiO*-polycrystalline | −1 | 0.24 | | 86/106ms | [33] |
| p-NiO/n-ZnO | PLD | *ZnO*- Crystalline *NiO*-Crystalline | 0 | 5m | $3.6 \times 10^{11}$ | 444/628μs | **This work** |

It is worth noting that although the rise time for the pyro current component is 3.92/8.9μs, the entire process, encompassing spike generation (pyro + photocurrent) and stabilization to a steady value (only photocurrent), takes approximately 2ms at zero bias condition. Our device does not show pyro-phototronic effect. In terms of response time, our device thus performs better than most of this type of photodetectors reported in the literature while offering quite comparable responsivity. Moreover, it should be noted that all the devices listed in Table 1 are based on polycrystalline NiO films. While our device is epitaxial heterojunction, which could be the reason for the faster response time. The epitaxial quality could also be the origin of the high level of reliability/stability of the device.

## CONCLUSION:

It has been found that epitaxial p-(001)NiO/n-(0001)ZnO heterostructures can be grown on *c*-sapphire substrates using PLD technique. These hetero-pn-junctions can act as self-powered UV-photodetectors. The device shows a peak responsivity of 5mA/W with fast rise (444μs) and decay (628μs) times could be achieved at zero bias condition. Interestingly, through applying bias, the photo-response time can be varied from several microseconds to thousands of seconds. This effect can be explained in terms of the field induced change in the capture barrier height associated with certain traps located at the junction, which governs the persistent photoconductivity effect. This feature can be exploited for neuromorphic device applications.

## SUPPORTING INFORMATION

See supporting information (S1) for I-V characteristics of NiO-NiO contact and ZnO-ZnO contact. (S2) Estimation of $\tau_d$ in reverse and forward bias.

## ACKNOWLEDGEMENTS

We acknowledge the financial support provided by Department of Science and Technology (DST) under Grant No: CRG/2022/001852, Government of India. Ms. Amandeep Kaur would like to thank Council of scientific

## AUTHOR DECLARATIONS

**Conflict of Interest**

The authors have no conflicts to disclose.

**Data Availability**

The data that support the findings of this study are available from the corresponding author on reasonable request

# p-(001)NiO/n-(0001)ZnO Heterostructures based Ultraviolet Photodetectors

Amandeep Kaur, Bhabani Prasad Sahu, Ajoy Biswas, Subhabrata Dhar*

Department of Physics, Indian Institute of Technology Bombay, Powai, Mumbai 400076, India

*Email: dhar@phy.iitb.ac.in

## S1: I-V characteristics of the contacts on NiO and ZnO sides

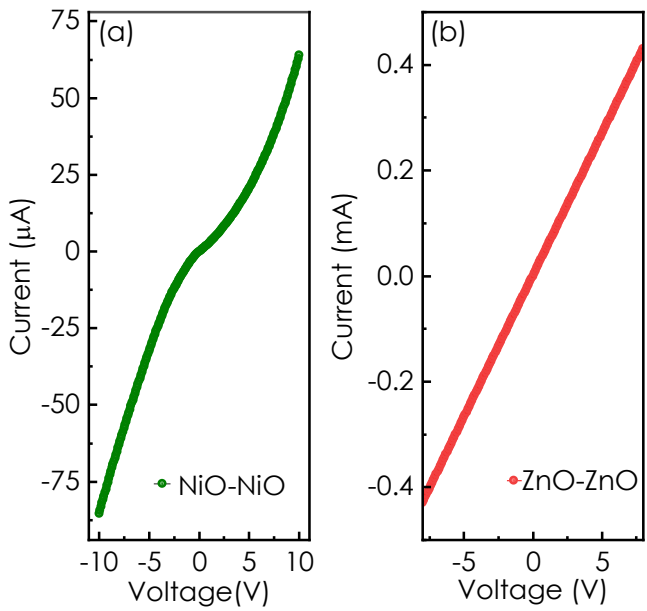

Fig. S1 Current-Voltage characteristics of the contacts on (a) NiO and (b) ZnO sides, showing nearly ohmic behaviour on the NiO side and ohmic behaviour on the ZnO side.

# S2: Estimation of $\tau_d$ from the slope of the log-of-current versus time plots

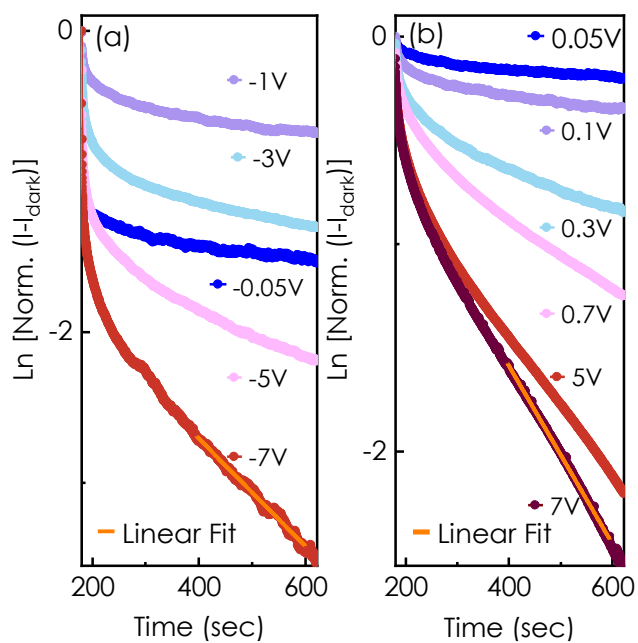

Fig S2 Log plots of the normalized photo-response profiles under different (a) reverse and (b) forward bias conditions.